\documentstyle[aps,prl,preprint,floats,epsfig]{revtex}  
\begin{document}
%\doubles

\preprint{\vbox{ \hbox{ BELLE-CONF-0138}}}
%\twocolumn[\hsize\textwidth\columnwidth\hsize\csname 
%@twocolumnfalse\endcsname 

\title{
 \quad\\[1cm] \Large
Observation of $B^+ \to \chi_{c0} K^+ $}

\author{The Belle Collaboration}

\maketitle
% to make it single spaced
\tighten

\begin{abstract}

  We report the first observation of the decay $B^+ \to \chi_{c0} K^+$
using 21.3~fb$^{-1}$ of data collected by the Belle detector at the 
$\Upsilon$(4S) resonance. The preliminary result for the branching fraction is
${\cal{B}}(B^+\to \chi_{c0}K^+)=(8.0^{+2.7}_{-2.4}\pm1.0\pm1.1)\times10^{-4}$
where the first error is  statistical, the second is systematic, and the third
comes from the uncertainty in the $\chi_{c0}\to\pi^+\pi^-$ branching fraction.

\end{abstract}

{\renewcommand{\thefootnote}{\fnsymbol{footnote}}

%% >>>>>> LP2001 authorlist will go here
%% >>>>>> current version is /g8home/browder/lp2001/authorlist_lp2001.tex

\begin{center}
  K.~Abe$^{9}$,               % KEK
  K.~Abe$^{37}$,              % TohokuGakuin
  R.~Abe$^{27}$,              % Niigata
  I.~Adachi$^{9}$,            % KEK
  Byoung~Sup~Ahn$^{16}$,      % Korea
  H.~Aihara$^{39}$,           % Tokyo
  M.~Akatsu$^{20}$,           % Nagoya
  K.~Asai$^{21}$,             % Nara
  M.~Asai$^{10}$,             % Hiroshima
  Y.~Asano$^{44}$
% ...
\end{center}

\small
\begin{center}
$^{1}${Aomori University, Aomori}\\
% ...
$^{9}${High Energy Accelerator Research Organization (KEK), Tsukuba}\\
$^{10}${Hiroshima Institute of Technology, Hiroshima}\\
% ...
$^{13}${Institute of High Energy Physics, Vienna}\\
% ...
$^{16}${Korea University, Seoul}\\
% ...
$^{20}${Nagoya University, Nagoya}\\
$^{21}${Nara Women's University, Nara}\\
% ...
$^{27}${Niigata University, Niigata}\\
% ...
$^{37}${Tohoku Gakuin University, Tagajo}\\
% ...
$^{39}${University of Tokyo, Tokyo}\\
% ...
$^{44}${University of Tsukuba, Tsukuba}\\
% ...
\end{center}

\normalsize

\setcounter{footnote}{0}
\newpage

\normalsize

%\onecolumn

\vspace*{0.5cm}

\section{Introduction}

   The production rate of quarkonium states in various high energy 
physics processes can provide valuable insight not only into the interactions 
between a heavy quark and antiquark, but also into the elementary 
processes with the  production of a  $Q\bar{Q}$ pair. 
The spin-1 $S$-wave resonances, such as the $J/\psi$ of the charmonium
family, are of special experimental significance because of their importance
for studies of $CP$ violation.
The $P$ states are also important in their own right because they probe a 
qualitatively different aspect of the $Q\bar{Q}$ production process. 
While the $S$ states probe only 
the production of color-singlet $Q\bar{Q}$ pairs at small separation
distances, the $P$ states can also probe the production of 
color-octet $Q\bar{Q}$ pairs. Color-singlet production of $\chi_{c0}$ 
in $B$ decays vanishes in the factorization 
approximation as a consequence of spin-parity conservation. 
However, the color-octet
mechanism allows for the production of the $\chi_{c0}$ $P$-wave $0^{++}$ state 
via the emission of a soft gluon~\cite{bodwin,beneke}. 

   The only recent search for the $B\to \chi_{c0} K$ decay was 
reported by CLEO~\cite{cleo}. Using $9.66 \times 10^6$~$B\bar{B}$ pairs
they set  90\% C.L. upper limits of: 
${\cal{B}}(B^+\to \chi_{c0}K^+) < 4.8\times10^{-4}$ and
${\cal{B}}(B^0\to \chi_{c0}K^0) < 5.0\times10^{-4}$.
This analysis uses a data sample
collected with the Belle detector~\cite{NIM} at KEKB asymmetric 
energy $e^+e^-$ collider~\cite{KEKB}.
It consists of 21.3 fb$^{-1}$ taken at the $\Upsilon(4S)$
(which corresponds to about 22.8 million produced $B\bar{B}$ pairs) and 
2.3 fb$^{-1}$ taken 60 MeV below for continuum studies.

   The detailed analysis of the final state presented in this paper is
part of the general analysis of $B$-meson three-body decays 
reported in ref.~\cite{khh}.

The inclusion of charge conjugate states is implicit 
throughout this report unless explicitly stated otherwise.

%%%%%%%%%%%%%%%%%%%%%%%%%%%%%%%%%%%%%%%%%%%%%%%%%%%%%%%%%%%%%%%%%%%%%%

\section{Event selection}

  Charged tracks are required to satisfy a set of track quality cuts
based on the average hit residual and  impact parameters in both  
$r$-$\phi$ and $r$-$z$ planes. We require that the transverse momentum of the 
track be greater than 100 MeV/$c$ to reduce low momentum combinatorial 
background. For more details, see ref.~\cite{khh}.

  Hadron identification is accomplished using
the responses of the ACC and the TOF and $dE/dx$ measurements in the CDC.
The information from these three subsystems is combined in a single 
number using the likelihood method:
\[ {\cal{L}}(h) = {\cal{L}}^{ACC}(h)\times {\cal{L}}^{TOF}(h)\times{\cal{L}}^{CDC}(h),\]
where $h$ stands for the hadron type ($\pi$, $K$, $p$).
Charged particles are 
identified as  $K$'s or $\pi$'s by cutting on the likelihood ratio (PID): 
\[ PID(K) = \frac{{\cal{L}}(K)}{{\cal{L}}(K)+{\cal{L}}(\pi)};
PID(\pi) = \frac{{\cal{L}}(\pi)}{{\cal{L}}(\pi)+{\cal{L}}(K)} = 1 - PID(K) \]
The likelihood
ratio for kaon candidates is required to be greater than 0.5 and for
pion candidates less than 0.9.

  All charged tracks are also required to satisfy an
electron veto requirement that demands
that the electron likelihood is less than 0.95.
In addition, all charged kaon candidates are required to satisfy a
proton veto:
\[ PID(p) = \frac{{\cal{L}}(p)}{{\cal{L}}(p)+{\cal{L}}(K)}<0.95.\]
For more details see rf.~\cite{khh} and references therein.

   We reconstruct $\chi_{c0}$ candidates in the 
$\chi_{c0} \to \pi^+\pi^-$ and $\chi_{c0} \to K^+K^-$ decay modes.
   The $B^+\to\chi_{c0}K^+$ candidate events are identified by means of
the beam-constrained mass $M_{BC}$ and the  energy difference $\Delta E$:
\[ M_{BC} = \sqrt{s/4 - P_B^{*2}}~;~~~\Delta E = E_B^* - \sqrt{s}/2, \]
where $E_B^*$ and $P_B^*$ are the energy and three-momentum of the $B$ 
candidate in the $\Upsilon (4S)$ rest frame and $\sqrt{s}$ is the total 
energy.
 Subsequently, we refer to  the ``$B$ signal region,'' which is defined as: 
\begin{center}
$5.272<M_{BC}<5.289$ GeV/$c^2$;~~~$|\Delta E|<40$~MeV.
\end{center}

%%%%%%%%%%%%%%%%%%%%%%%%%%%%%%%%%%%%%%%%%%%%%%%%%%%%%%%%%%%%%%%%%%%%%%

\section{Background suppression}

   To suppress the combinatorial background which is dominated 
by the two-jet-like $e^+e^-\to~q\bar{q}$ continuum process, we use variables 
that characterize the event topology.
   We require $|\cos(\theta_{Thr})|<0.80$ where $\theta_{Thr}$ 
is the angle between the thrust axis of the $B$ candidate and that 
of the rest of the event. This eliminates 83\% of the continuum 
background and retains 79\% of the signal events. 
Following the analysis of ref.~\cite{khh}, we define a 
Fisher discriminant ${\cal F}$ that includes the angle of the thrust axis of
the entire event, the production angle of the $B$ meson candidate
and nine ``Virtual Calorimeter''~\cite{VCal} parameters that characterize the
momentum flow in the event relative to  the $B$ candidate thrust axis.
When the requirement on the Fisher discriminant 
variable  ${\cal{F}}>0.5$ is imposed,  79\% of the continuum background is
rejected with about 74\% efficiency for the signal. 
In the case of three charged kaon 
final states, the continuum background is much smaller and 
a looser requirement ${\cal{F}}>0$ can be imposed. 
This rejects 53\% of the continuum background with about 
89\% efficiency for the signal.

   In the $K^+\pi^+\pi^-$ final state the background from $B$ generic 
decays is dominated by the decays of
type $B^+\to [K^+\pi^-]\pi^+$ where $[K^+\pi^-]$ denotes a resonance state 
which can decay into $K^+\pi^-$ such as $\bar{D^0}$, $K^{*0}(892)$ etc.
To suppress this type of background, we require the invariant mass of 
the $K^+\pi^-$ system be greater than 2.0 GeV/c$^2$.
For the $K^+K^+K^-$ final state we require the invariant mass for 
both $K^+K^-$ combinations  be greater than 2.0 GeV/c$^2$ to suppress the 
possible background from rare $B$ decays such as $B^+\to \phi K^+$.

\section{Results of the Analysis }

\subsection{Signal yield extraction}

   First, we select all $K^+\pi^+\pi^-$ ($K^+K^+K^-$) 
combinations from the $B$ signal region that satisfy
the selection criteria described above and have the   
$\pi^+\pi^-$ ($K^+K^-$) invariant mass in the range
$3.2 < M(h^+h^-) < 3.8$~GeV/c$^2$. The resulting 
$\pi^+\pi^-$ and $K^+K^-$ invariant mass spectra
are shown in Figs.~\ref{hhmass}a and~\ref{hhmass}b, respectively.
Since in the case of the $K^+K^+K^-$ final state
there are two same charge kaons,
we distinguish the $K^+K^-$ combination with the smaller, $M(K^+K^-)_{min}$, 
and larger, $M(K^+K^-)_{max}$, invariant masses.
Only the larger combination is plotted.
Two prominent peaks can be seen in Fig.~\ref{hhmass}a.
The peak around 3.69~GeV/c$^2$ corresponds to
the $\psi(2S)$ meson from the decay mode
$B^+\to \psi(2S)K^+$, $\psi(2S)\to \mu^+\mu^-$ with muons misidentified 
as pions. It is slightly shifted to higher mass values 
due to the wrong mass assignment for $\psi(2S)$ daughter particles.
The peak just above 3.4 GeV/c$^2$ is identified as the $\chi_{c0}$ meson.

   The $\pi^+\pi^-$ and $K^+K^-$ spectra are fitted to the sum of a 
Breit-Wigner function convolved with a Gaussian
for the signal and a zeroth-order polynomial for the background.
The width of the Gaussian  is fixed at $10.8$ MeV/$c^2$
as determined from a fit to the $J/\psi$ peak in the $\mu^+\mu^-$ 
invariant mass spectrum.
The width of the Breit-Wigner function is fixed at the 
world average value for the $\chi_{c0}$~\cite{PDG}.
The results of the fit to the $\pi^+\pi^-$ and $K^+K^-$ invariant mass 
spectra are given in Table~\ref{hhfit}.
The peak position in the $K^+K^-$ spectrum, Fig.~\ref{hhmass}b, 
is found to be somewhat shifted to lower mass values.
Although this shift is consistent with a statistical fluctuation,
we note that according to the results of our general analysis
of the three body $B^+\to K^+K^+K^-$ decay, we find a significant signal
in the mass sidebands of the $\chi_{c0}\to K^+K^-$ signal~\cite{khh}. 
As a result, the $K^+K^-$
invariant mass distribution in the $\chi_{c0}$ region could be distorted
by the effects of interference with an amplitude not related to the
$B^+\to\chi_{c0}K^+$.  A Monte Carlo simulation study indicates that
this effect could shift the $\chi_{c0}$ peak position 
by as much as  $15$ MeV/$c^2$ and 
could have as much as a 100\% effect on the observed amplitude
of the $\chi_{c0}$ signal.
Because of this uncertainty we base our branching fraction measurement for
the $B^+\to\chi_{c0}K^+$ decay on the $\chi_{c0}\to \pi^+\pi^-$
decay mode only.

% ================================================================================ %
\begin{table}[hbt]
\caption{Results of the fit to the $\pi^+\pi^-$ and $K^+K^-$ invariant mass spectra.}
\medskip
\label{hhfit}
  \begin{tabular}{lcccr}
  $\chi_{c0}$ submode \hspace*{0.1cm} & Efficiency, \% 
                                      & Peak, GeV/$c^2$
                                      & Yield, events  
                                      & Significance, $\sigma$ \\ \hline
  $\chi_{c0}\to\pi^+\pi^-$ & $21.5$ & $3.408\pm0.006$ & $15.5^{+5.3}_{-4.6}$ & $4.8$ \\
  $\chi_{c0}\to K^+K^-$    & $13.7$ & $3.390\pm0.009$ & $7.7^{+3.9}_{-3.1}$ & $3.2$ \\
  \end{tabular}
\end{table}
%  =============================================================================== %

  The $\chi_{c0}$ candidates are then selected by requiring
$|M(h^+h^-)-3.415| < 0.050$~GeV/c$^2$, where $h^+h^-$ denotes both 
$\pi^+\pi^-$ or $K^+K^-$ combinations.
This corresponds to about a 2.6~$\sigma$ cut. The resulting two-dimensional
$\Delta E$ versus $M_{BC}$  plots as well as the projections on 
the $\Delta E$ and $M_{BC}$ axes are shown in
Figs.~\ref{mbdepp} and~\ref{mbdekk} for the
$K^+\pi^+\pi^-$ and $K^+K^+K^-$ final states, respectively.
Clear signals are apparent in both figures.

%%%%%%%%%%%%%%%%%%%%%%%%%%%%%%%%%%%%%%%%%%%%%%%%%%%%%%%%%%%%%%%%%%%%%%%%%%%%%%%%%%%

\subsection{Branching fraction calculation}

   To determine branching fractions, we 
normalize our results to the observed 
$B^+\to \bar{D}^0\pi^+$, $\bar{D}^0\to K^+\pi^-$ signal. 
Although this introduces a $9.7$\% systematic error 
because of the uncertainty in the
$B^+\to \bar{D}^0\pi^+$ branching fraction, it removes systematic
effects in the particle identification efficiency,
charged track reconstruction efficiency and the systematic uncertainty 
due to the cuts on event shape variables.
   We calculate the branching fraction for the $B$ meson decay to a 
final state $f$ via the relation:
\begin{equation}
   {\cal{B}}(B\to f) = 
   {\cal{B}}(B^+\to \bar{D}^0\pi^+)\times{\cal{B}}(\bar{D}^0\to K^+\pi^-)
   \frac{N_f}{N_{D\pi}}\times\frac{\varepsilon_{D\pi}}{\varepsilon_{f}},
\end{equation}
where $N_f$ and $N_{D\pi}$ are the numbers of 
observed events for the particular final state 
$f$ and for the reference process respectively,
$\varepsilon_{f}$ and $\varepsilon_{D\pi}$ are corresponding 
reconstruction efficiencies determined from the Monte Carlo simulation.

As a cross-check, we  use the decay mode
$B^+\to J/\psi K^+$ followed by the $J/\psi \to \mu^+\mu^-$ decay.
In this case there are
two muons in the final state instead of two pions.
To avoid additional systematic 
uncertainty in the muon identification efficiency,
we do not use muon identification information for $J/\psi$ 
reconstruction. Instead, we apply the same pion-kaon separation 
cut for muons from $J/\psi$ as for pions from  $\chi_{c0}$. 
The feed-down from the $J/\psi\to~e^+e^-$ submode is found to be 
negligible (less than 0.5\%) after the
application of the electron veto requirement.

   The signal yields for the $B^+\to \bar{D^0}\pi^+$ and $B^+\to J/\psi K^+$
processes are extracted from fits to the $\Delta E$ 
distributions shown in Fig.~\ref{deref}. The yields are
found to be  $N_{D\pi}=858\pm31$ and $N_{J/\psi}=323\pm19$ respectively.
Using the corresponding reconstruction efficiencies of $23.5$\% and $29.4$\%
we find the ratio, $R$, of branching fractions to be 
\[ R = \frac{{\cal{B}}(B^+\to J/\psi K^+)}{{\cal{B}}(B^+\to \bar{D}^0\pi^+)} = 
\frac{N_{J/\psi K}}{N_{D\pi}}\times
\frac{\varepsilon_{D\pi}}{\varepsilon_{J/\psi K}}\times
\frac{{\cal{B}}(\bar{D}^0\to K^+\pi^-)}{{\cal{B}}(J/\psi\to\mu^+\mu^-)} 
= 0.196\pm0.014,\]
%\pm0.006,\]
%
where only the statistical error is quoted.
%where the first error is statistical and the second is due to 
%uncertainty in the secondary branching fractions. 
This result is in good agreement with the value calculated 
from PDG~\cite{PDG} data: $0.189\pm0.026$.

   To calculate the branching fraction for 
the $B^+\to \chi_{c0}K^+$ decay mode,
we use the signal yield determined from the fit to the $h^+h^-$ 
invariant mass spectra to take into account the 
possible contribution from the 
non-resonant $B^+\to K^+h^+h^-$ decays that would produce signal-like 
distributions in both $\Delta E$ and $M_{BC}.$
Combining all the relevant numbers from Table~\ref{hhfit}
and the intermediate branching fractions 
from PDG~\cite{PDG}, we find the branching fraction for 
the $B^+\to \chi_{c0}K^+$ decay mode to be:
\begin{center}
 ${\cal{B}}(B^+\to \chi_{c0}K^+)=(8.0^{+2.8}_{-2.4}\pm1.0\pm1.1)\times10^{-4}$
 ~~($\chi_{c0} \to \pi^+\pi^-$ mode);
\vspace*{0.3cm}\\
 ${\cal{B}}(B^+\to \chi_{c0}K^+) =(5.3^{+2.7}_{-2.2}\pm0.7\pm0.8)\times10^{-4}$
 ~~($\chi_{c0} \to K^+K^-$ mode),
\end{center}
where the first error is statistical, the second is systematic,
%due to the uncertainty in 
%the $B^+\to \bar{D^0}\pi^+$ and $\bar{D^0}\to K^+\pi^-$
%branching fractions, 
and the third is due to the uncertainty in the 
$\chi_{c0}\to h^+h^-$ branching fractions.
The systematic error consists of the uncertainty in the 
$B^+\to \bar{D^0}\pi^+$ and $\bar{D^0}\to K^+\pi^-$ branching fractions (9.7\%),
the uncertainty in the background parameterization in the fit to the 
$h^+h^-$ spectra (7.8\% for $\pi^+\pi^-$ and 8.5\% for $K^+K^-$) and
the uncertainty in the background and signal parameterization in the fit to the
$\Delta E$ distribution for $B^+\to \bar{D^0}\pi^+$ signal (2.3\%).
%We include the second error which comes from the uncertainty in the
%branching fractions for the reference process in the 
%systematic error.
As mentioned above, our qualitative estimates indicate that
the model dependent error in the $K^+K^+K^-$ final state could be
very large. We do not present the quantitative estimation here because 
of the large uncertainty.

   We also calculate the ratio of the branching fractions for the
$B^+\to\chi_{c0}K^+$ and
$B^+\to~J/\psi K^+$ decays using only the $\chi_{c0}\to \pi^+\pi^-$ submode:
\[ \frac{{\cal{B}}(B^+\to \chi_{c0}K^+)}{{\cal{B}}(B^+\to J/\psi K^+)} = 
0.77^{+0.27}_{-0.23}\pm0.11, \]
where the first error is statistical and the second comes from the 
uncertainties in the \mbox{$J/\psi\to\mu^+\mu^-$} and $\chi_{c0}\to \pi^+\pi^-$ 
branching fractions.

%%%%%%%%%%%%%%%%%%%%%%%%%%%%%%%%%%%%%%%%%%%%%%%%%%%%%%%%%%%%%%%%%%%%%%%%%%%%%%%%%%%

\subsection{Cross-check with $\chi_{c0}\to K^{*0}K^-\pi^+$ mode}

   As an additional cross-check, we reconstruct the
$\chi_{c0}$ meson in the multi-body
$K^+K^-\pi^+\pi^-$ final state. Although, in general,  
the branching fractions for $\chi_{c0}$ decays to multi-body final
states are significantly higher than those for $\pi^+\pi^-$ and 
$K^+K^-$, these modes suffer from a much larger combinatorial background.
This background is somewhat reduced
in the $K^+K^-\pi^+\pi^-$ final state because of the presence of two charged
kaons. We further reduce the
combinatorial background by requiring that at least one $K\pi$ pair be 
consistent with $K^{*}(892)\to K\pi$: 
$|M(K\pi)-0.896|<0.050$~GeV/$c^2$.
We also use a tighter cut on the Fisher discriminant variable 
${\cal{F}}>0.8$ to suppress the larger continuum background.

 The resulting two-dimensional $\Delta E$ versus $M_{BC}$ plot as well as the
$\Delta E$ and $M_{BC}$ projections are presented in Fig.~\ref{mbdekkpp}.
From the fit to the $\Delta E$ distribution $9.2^{+3.9}_{-3.4}$ signal
events are found. Using the
reconstruction efficiency of $7.2$\% from Monte Carlo,
we obtain a branching fraction
that is in agreement with that determined for the $\pi^+\pi^-$ mode, namely
\begin{center}
 ${\cal{B}}(B^+\to \chi_{c0}K^+)=(7.0^{+3.0}_{-2.6}\pm2.3)\times10^{-4}$
 ~~($\chi_{c0} \to K^{*0}K^-\pi^+$ mode),
\end{center}
where the first error is statistical and the second comes from the 
uncertainty in the $\chi_{c0}\to~K^{*0}K^-\pi^+$ branching fraction.

\section{Conclusion}

  We report the first observation of the $B^+\to\chi_{c0}K^+$ decay
mode. The 
statistical significance of the signal is more than $4\sigma$. 
The preliminary branching fraction result is 
${\cal{B}}(B^+\to \chi_{c0}K^+)=(8.0^{+2.7}_{-2.4}\pm1.0\pm1.1)\times10^{-4}$
which is comparable to that for the $B^+\to J/\psi K^+$ decay. 
This provides evidence for a significant nonfactorizable contribution
% from a color-octet mechanism 
in $B$ to charmonium decay processes.

  The branching fraction measured by Belle is slightly higher than the CLEO 
published upper limit. However, given the uncertainties
on the result, it would be premature to claim that they disagree.

\section*{Acknowledgement}

%% Please paste this acknowledgement into your latex file.  

We wish to thank the KEKB accelerator group for the excellent
operation of the KEKB accelerator. We acknowledge support from the
Ministry of Education, Culture, Sports, Science, and Technology of Japan
and the Japan Society for the Promotion of Science; the Australian
Research
Council and the Australian Department of Industry, Science and
Resources; the Department of Science and Technology of India; the BK21
program of the Ministry of Education of Korea and the CHEP SRC
program of the Korea Science and Engineering Foundation; the Polish
State Committee for Scientific Research under contract No.2P03B 17017;
the Ministry of Science and Technology of Russian Federation; the
National Science Council and the Ministry of Education of Taiwan; the
Japan-Taiwan Cooperative Program of the Interchange Association; and
the U.S. Department of Energy.

\newpage

\begin{figure}[htb]
  \centering
    \includegraphics[height=10.0cm,width=12.0cm]{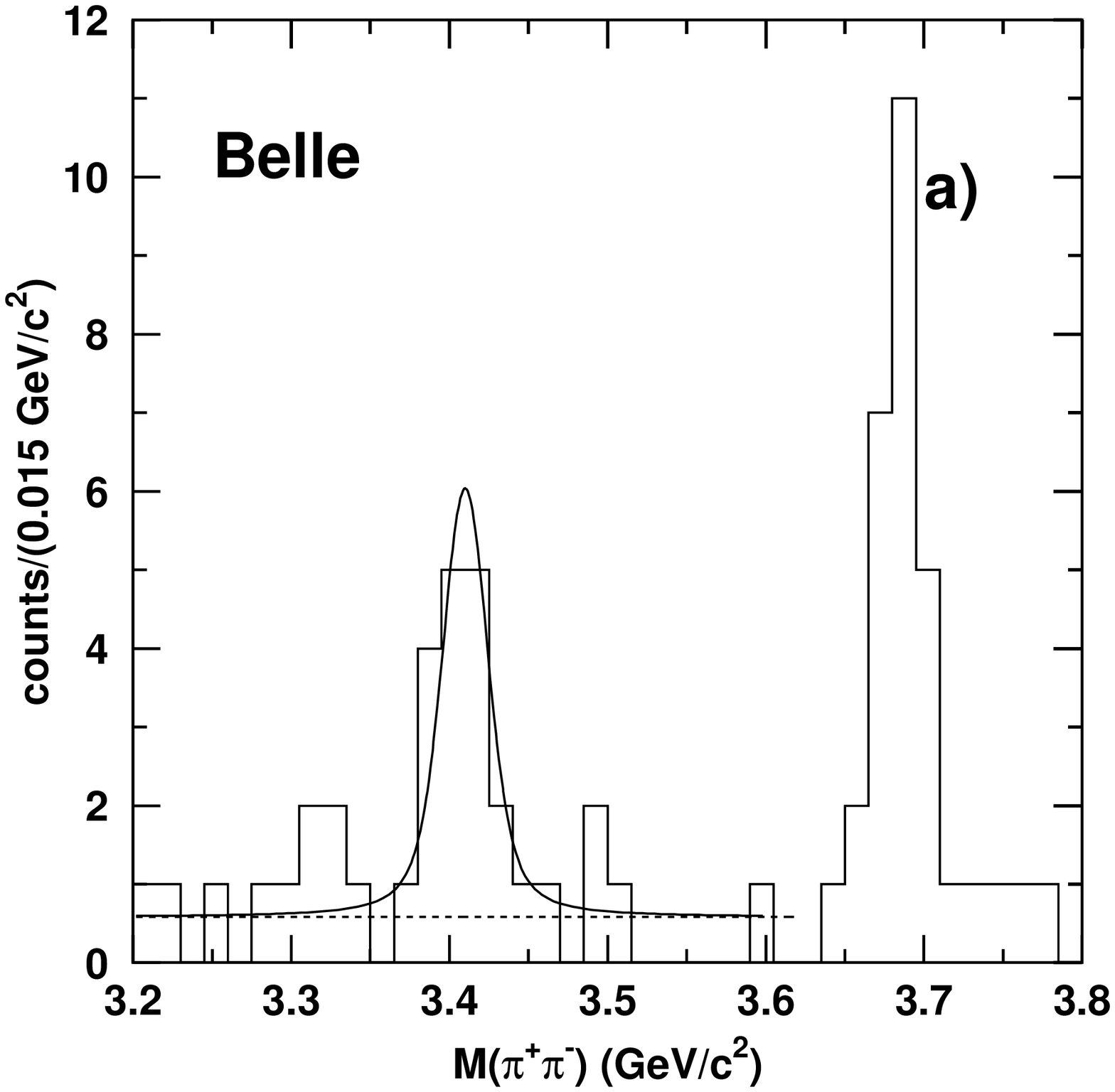} \hfill
    \includegraphics[height=10.0cm,width=12.0cm]{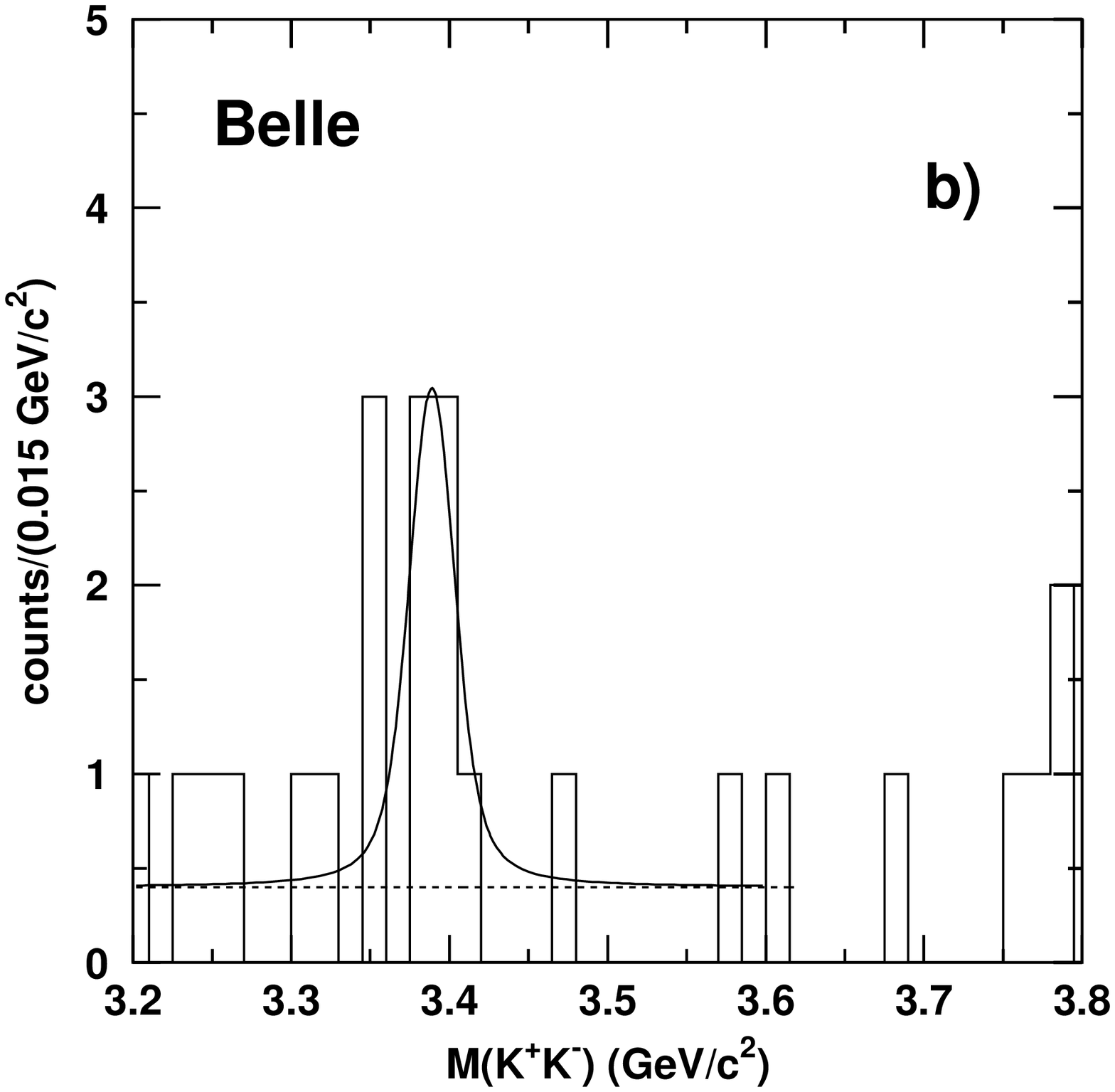}
    \\
  \centering
    \caption{The $\pi^+\pi^-$ (a) and the $K^+K^-$ (b)
             invariant mass spectra for $B$ candidates from the signal region.
             The histograms are data and the curves are fits to the data.}
    \label{hhmass}
\end{figure}

%%%%%%%%%%%%%%%%%%%%%%%%%%%%%%%%%%%%%%%%%%%%%%%%%%%%%%%%%%%%%%%%%%%%%%%%%%%%%%%%%%%%

\begin{figure}[htb]
  \includegraphics[height=8.0cm,width=8.0cm]{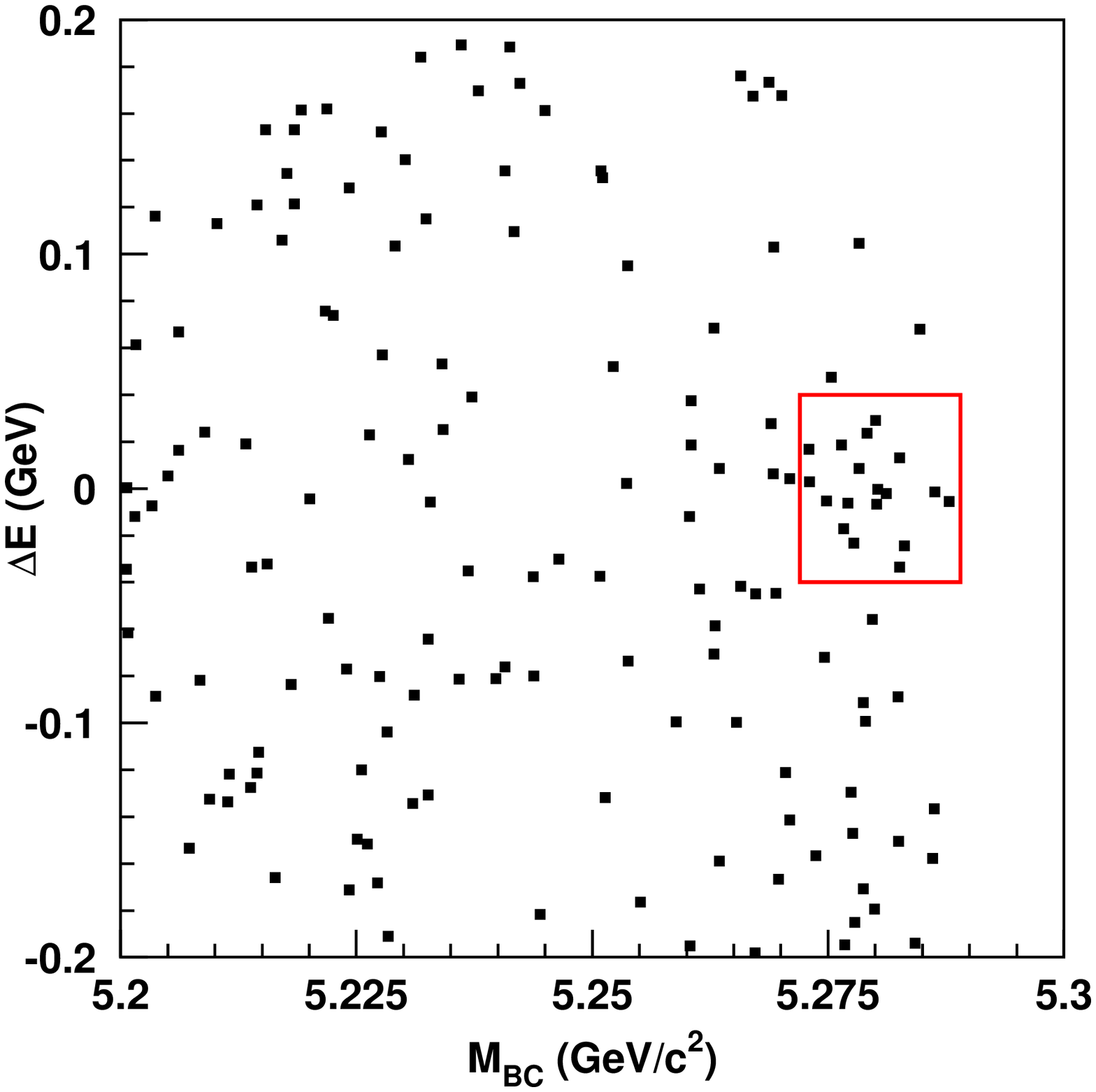} \hfill
  \includegraphics[height=8.0cm,width=8.0cm]{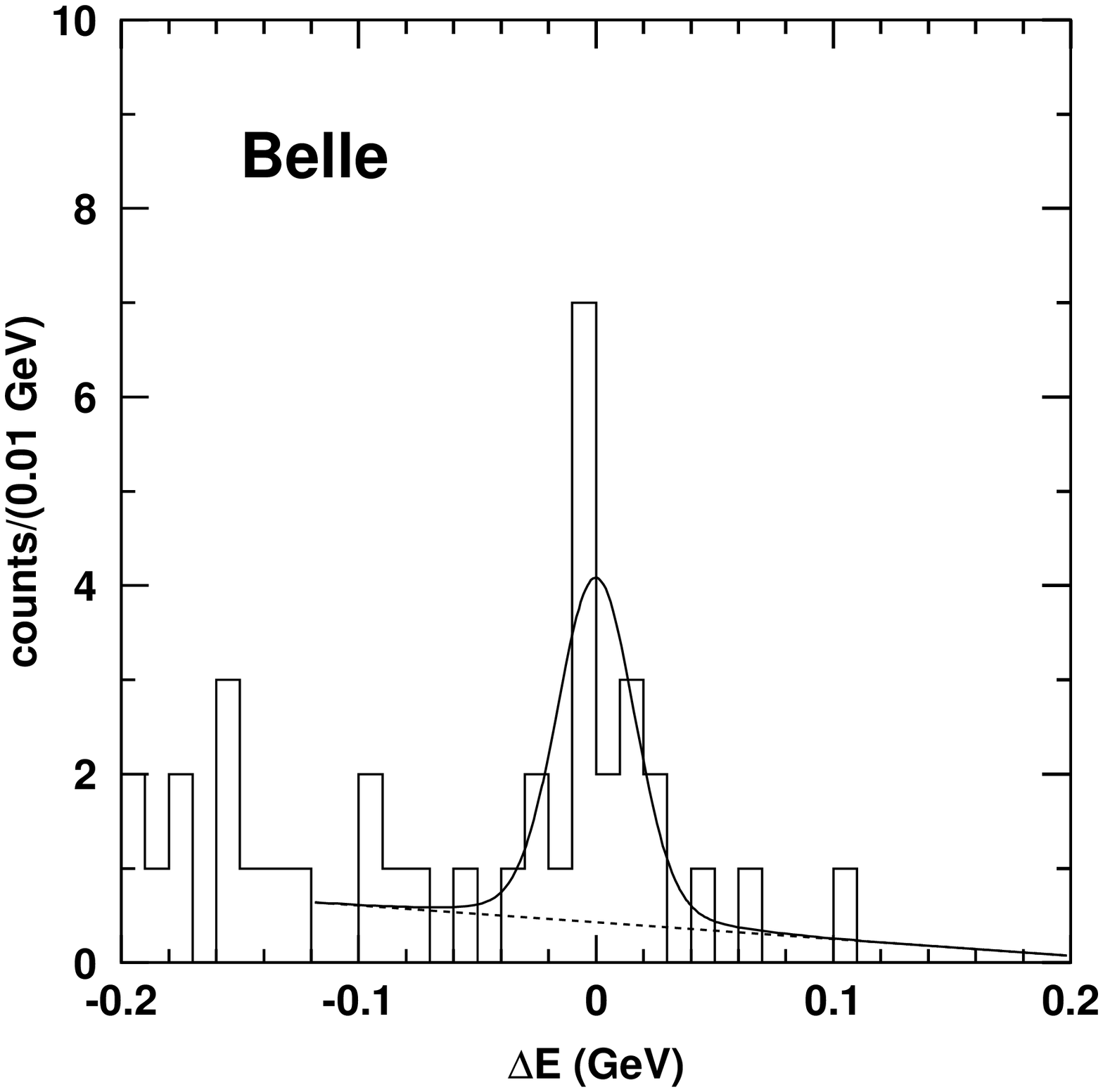} \\
  \includegraphics[height=8.0cm,width=8.0cm]{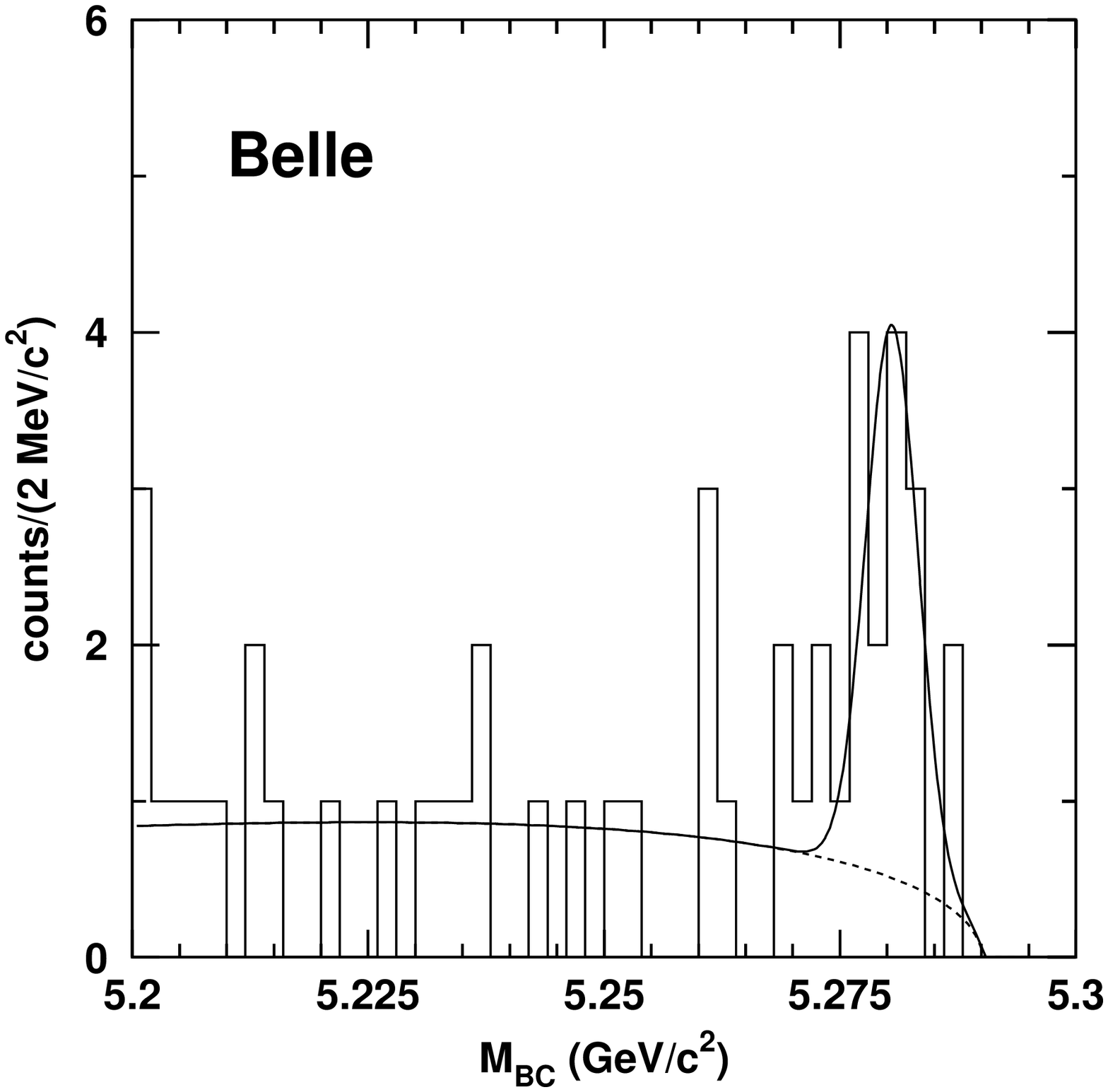} \hfill
%    \hspace*{-1.0cm}\includegraphics[height=5.5cm,width=8.5cm]{plots/dummy.eps} 
  \caption{The $\Delta E$ and $M_{bc}$ distributions for $B^+ \to \chi_{c0}K^+$,
           $\chi_{c0}\to\pi^+\pi^-$ candidates.
           The box in the two-dimensional plot shows the signal region.}
  \label{mbdepp}
\end{figure}

%%%%%%%%%%%%%%%%%%%%%%%%%%%%%%%%%%%%%%%%%%%%%%%%%%%%%%%%%%%%%%%%%%%%%%%%%%%%%%%%%%%%

\begin{figure}[htb]
  \includegraphics[height=8.0cm,width=8.0cm]{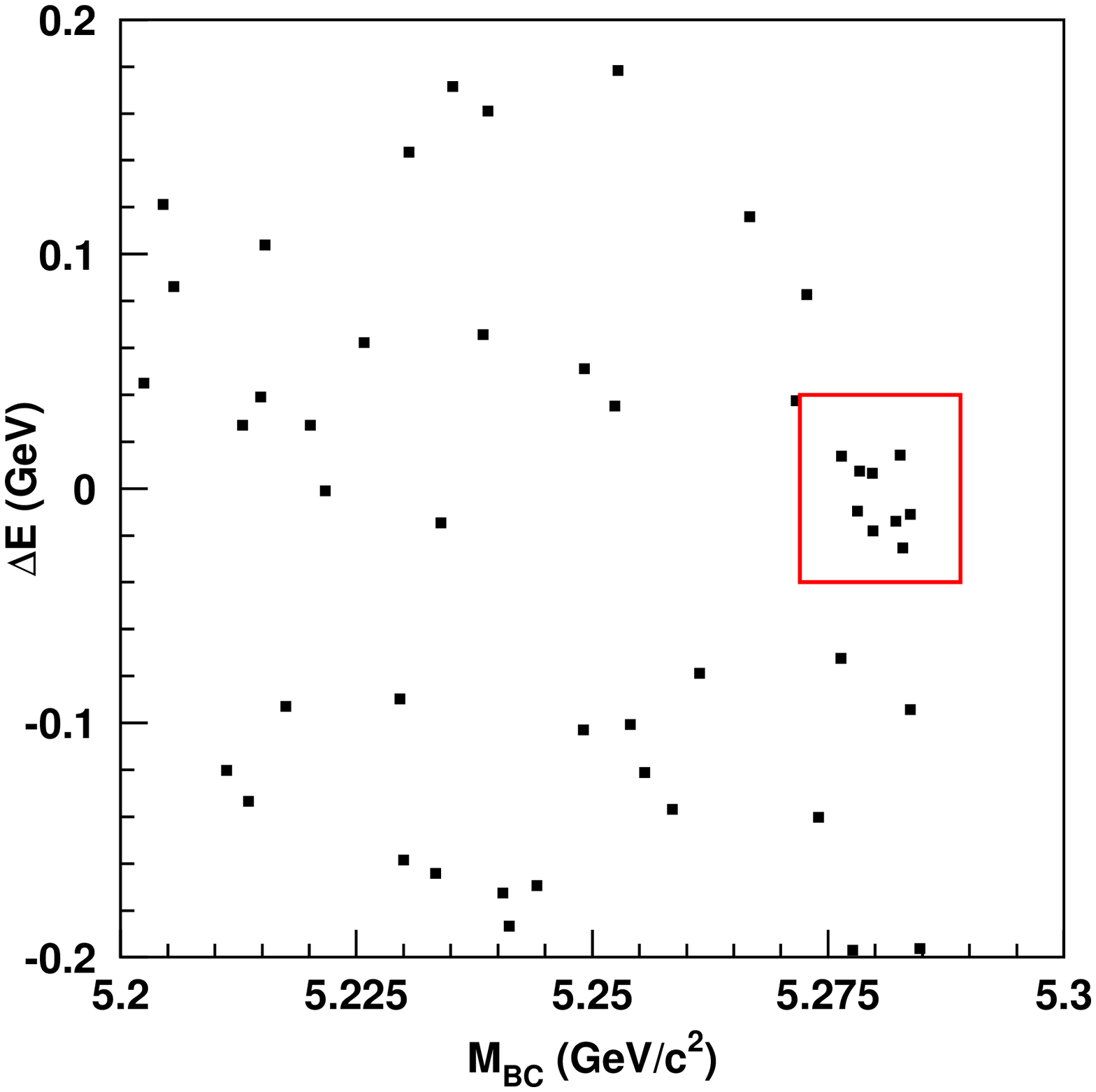} \hfill
  \includegraphics[height=8.0cm,width=8.0cm]{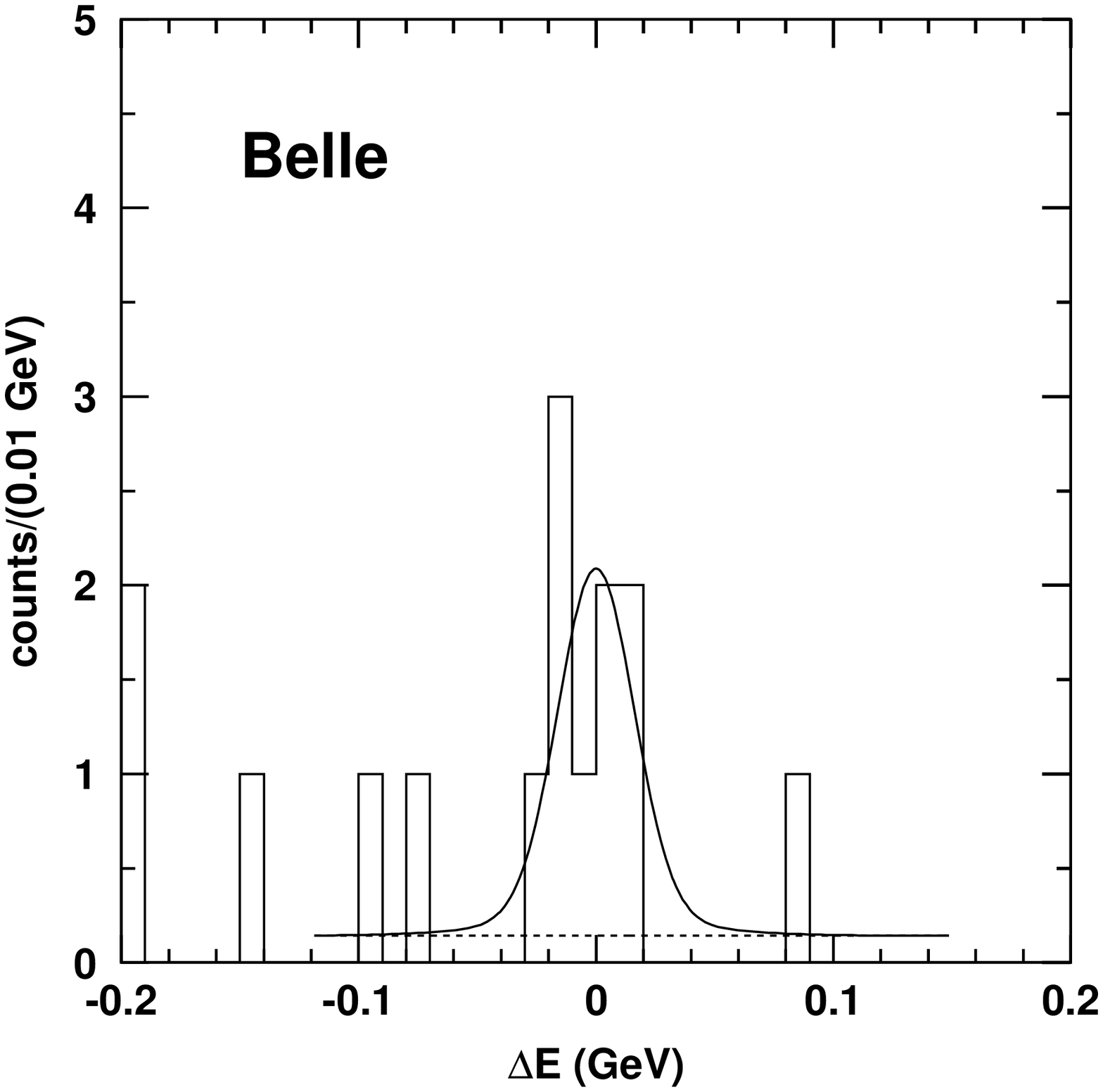} \\
  \includegraphics[height=8.0cm,width=8.0cm]{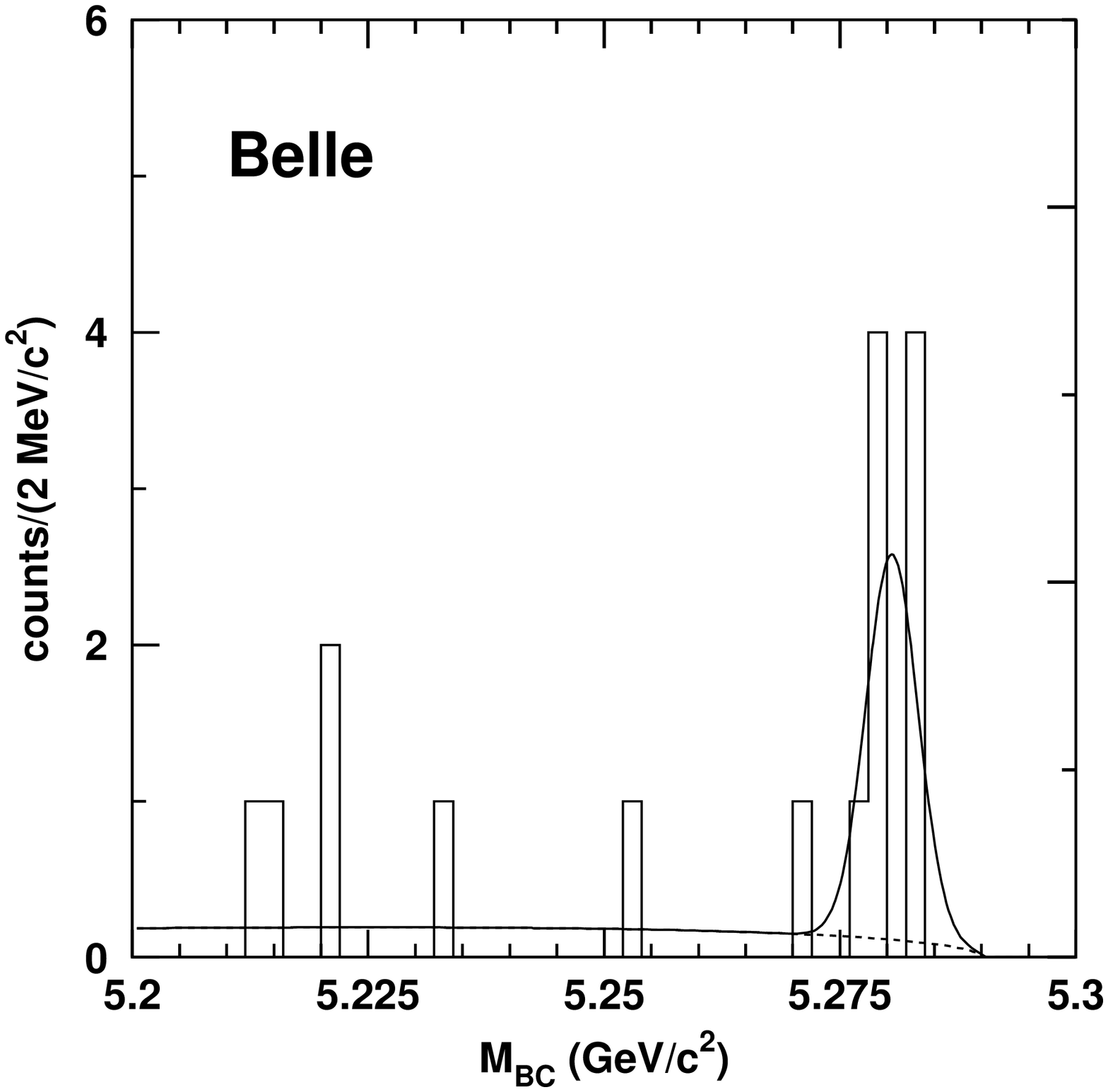} \hfill
%    \hspace*{-1.0cm}\includegraphics[height=5.5cm,width=8.5cm]{plots/dummy.eps} 
  \caption{The $\Delta E$ and $M_{bc}$ distributions for $B^+ \to \chi_{c0}K^+$,
           $\chi_{c0}\to K^+K^-$ candidates.
           The box in the two-dimensional plot shows the signal region.}
  \label{mbdekk}
\end{figure}

%%%%%%%%%%%%%%%%%%%%%%%%%%%%%%%%%%%%%%%%%%%%%%%%%%%%%%%%%%%%%%%%%%%%%%%%%%%%%%%%%%%%

\begin{figure}[thb]
  \centering
    \includegraphics[height=10.0cm,width=12.0cm]{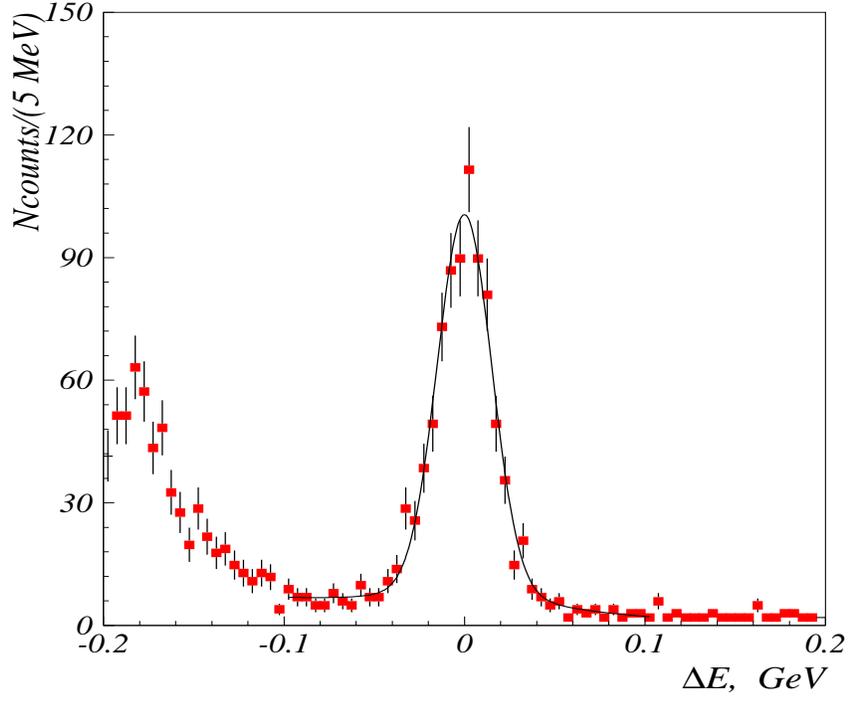}
    \includegraphics[height=10.0cm,width=12.0cm]{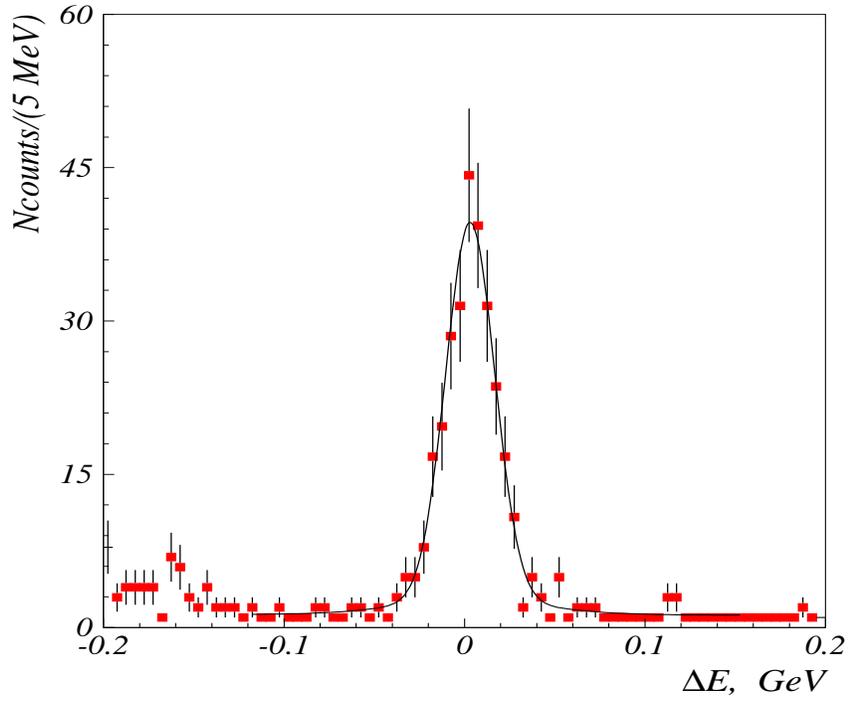}
  \centering
  \caption{The $\Delta E$ distributions for reference processes: 
           {\bf a)}~-~$B^+\to \bar{D^o}\pi^+$, $\bar{D^o}\to K^+\pi^-$;
           {\bf b)}~-~$B^+\to J/\psi K^+$, $J/\psi\to \mu^+\mu^-$.}
  \label{deref}
\end{figure}

%%%%%%%%%%%%%%%%%%%%%%%%%%%%%%%%%%%%%%%%%%%%%%%%%%%%%%%%%%%%%%%%%%%%%%%%%%%%%%%%%%%%%

\begin{figure}[htb]
  \includegraphics[height=8.0cm,width=8.0cm]{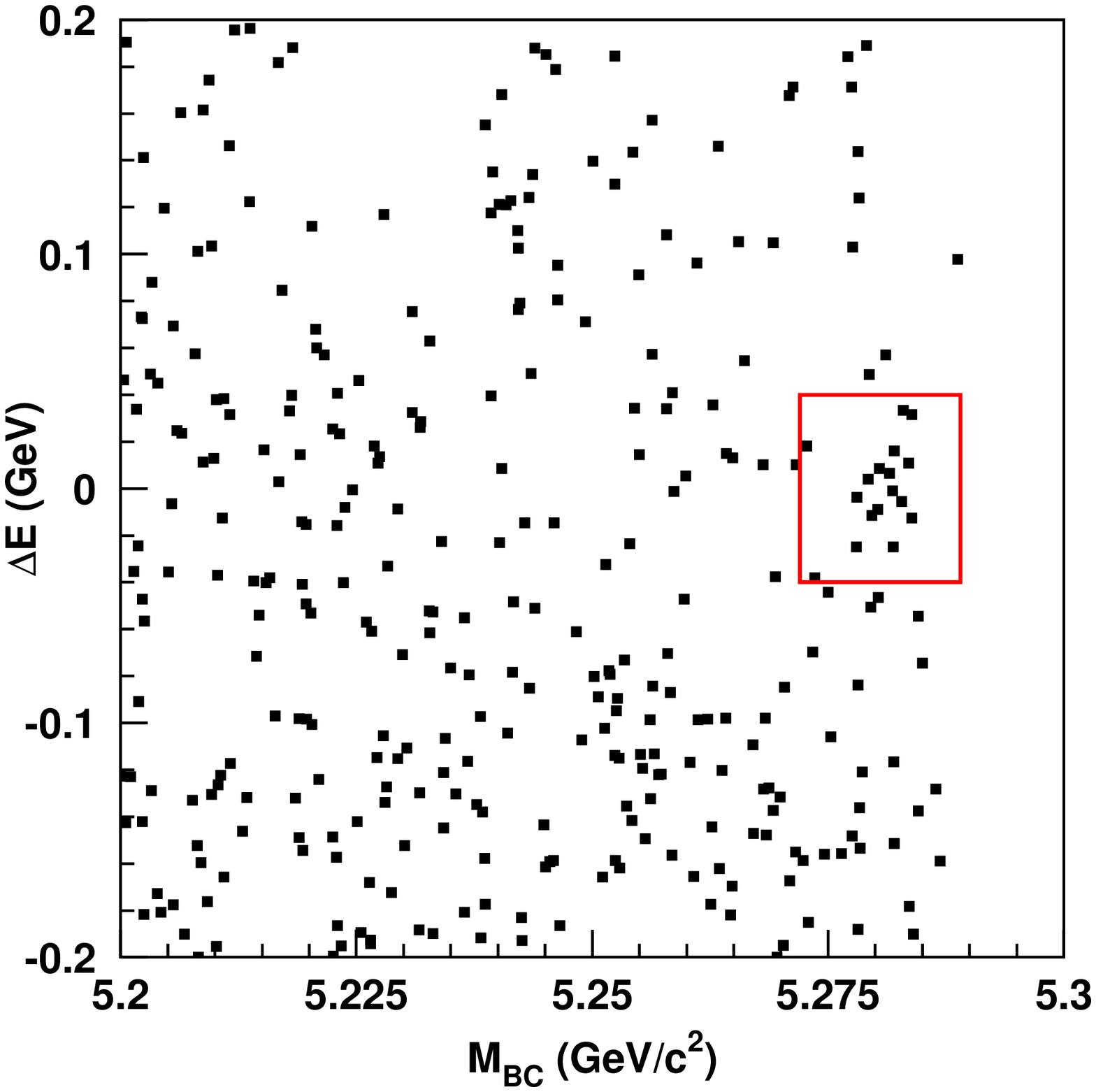} \hfill
  \includegraphics[height=8.0cm,width=8.0cm]{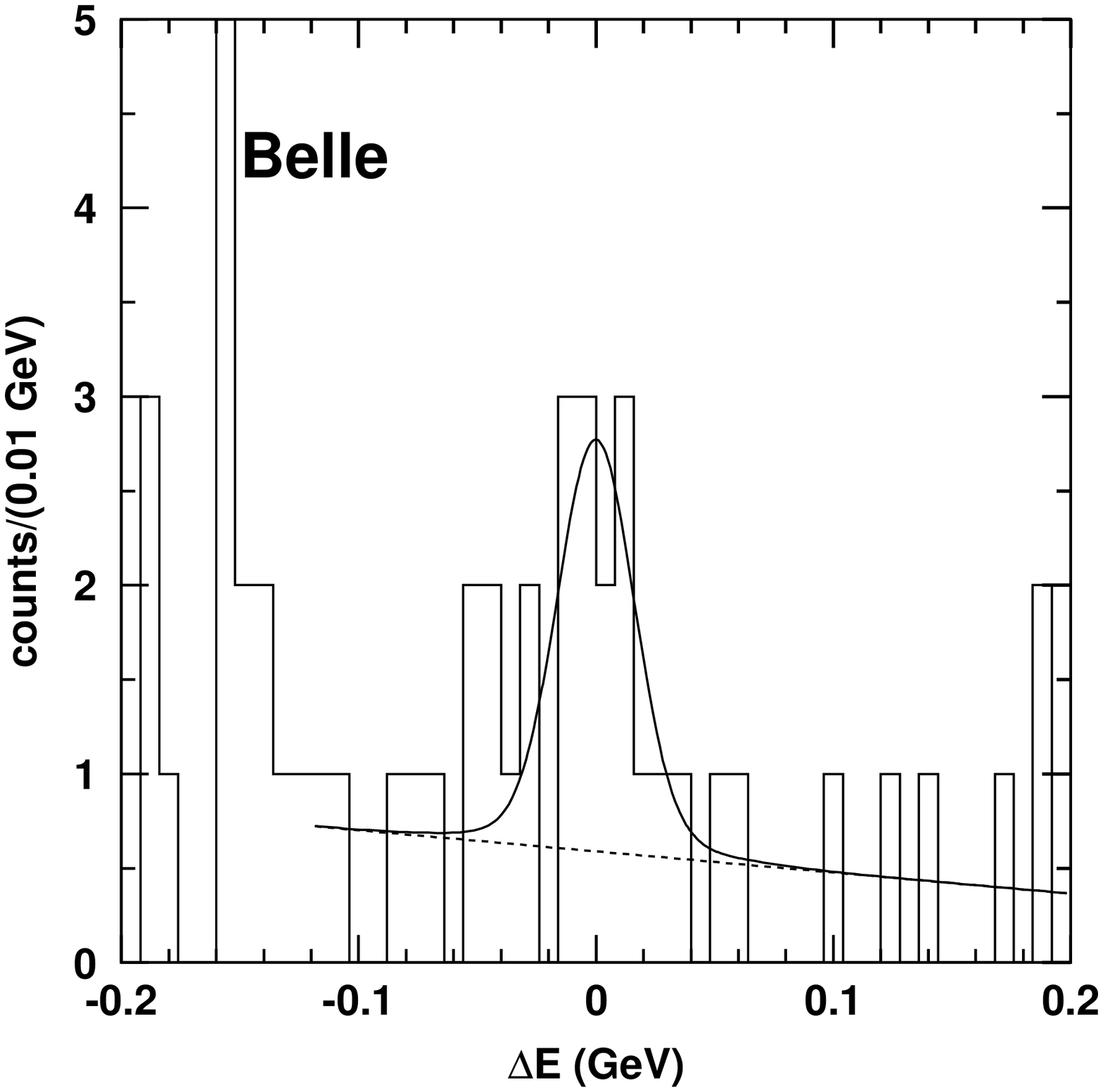} \\
  \includegraphics[height=8.0cm,width=8.0cm]{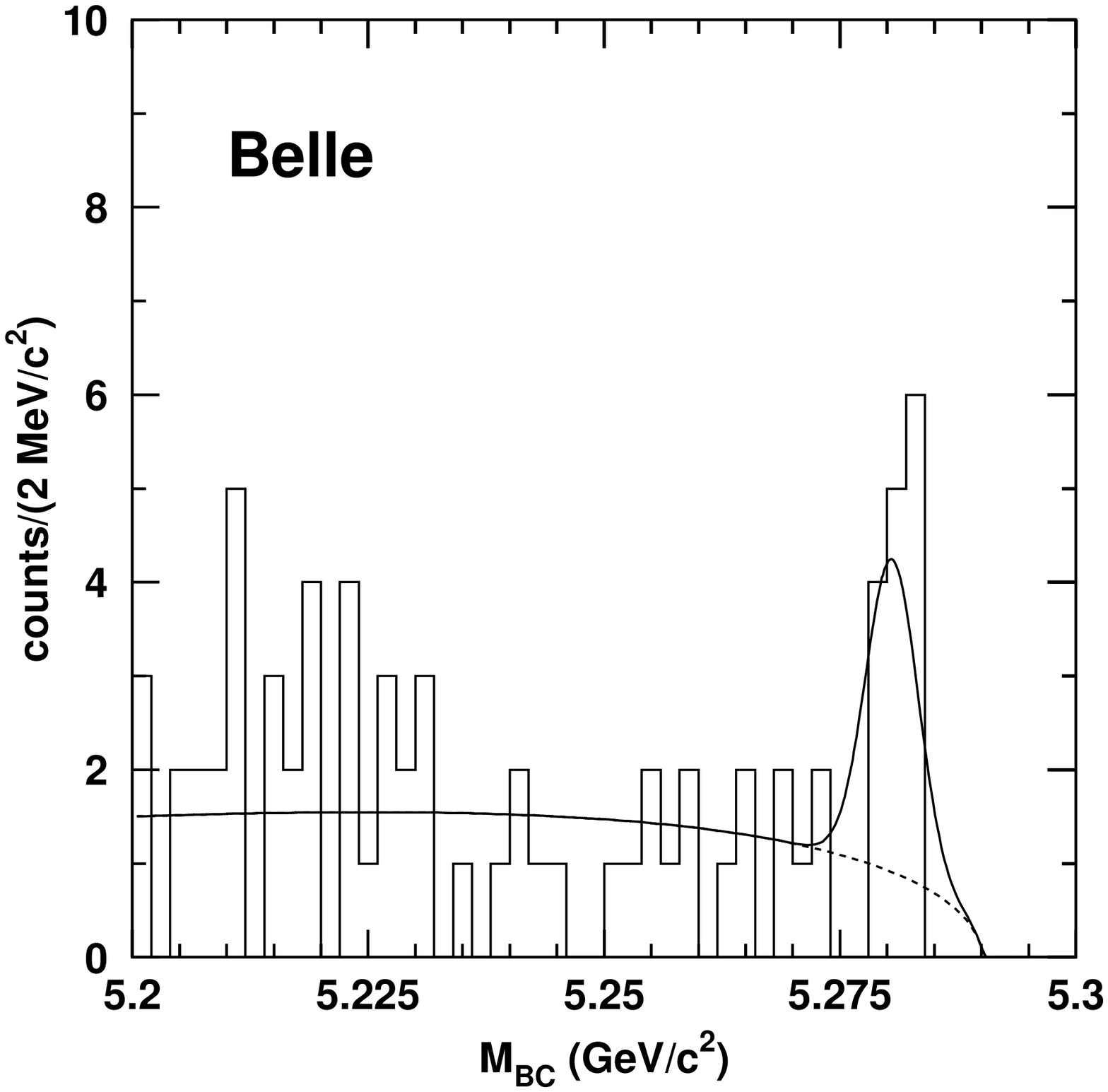} \hfill
%    \hspace*{-1.0cm}\includegraphics[height=5.5cm,width=8.5cm]{plots/dummy.eps} 
  \caption{The $\Delta E$ and $M_{bc}$ distributions for $B^+ \to \chi_{c0}K^+$,
           $\chi_{c0}\to K^{*o}K^+\pi^-$ candidates.
           The box in the two dimensional plot shows the signal region.}
  \label{mbdekkpp}
\end{figure}

%%%%%%%%%%%%%%%%%%%%%%%%%%%%%%%%%%%%%%%%%%%%%%%%%%%%%%%%%%%%%%%%%%%%%%%%%%%%%%%%%%%%

%%%%%%%%%%%%%%%%%%%%%%%%%%%%%%%%%%%%%%%%%%%%%%%%%%%%%%%%%%%%%%%%%%%%%%%%%%%%%%%

\begin{thebibliography}{99}


\bibitem{bodwin}{G.~Bodwin {\it et al.}, Phys. Rev. D {\bf46}, 3703 (1992).}
%                
\bibitem{beneke}{M.~Beneke {\it et al.}, Phys. Rev. D {\bf59}, 054003 (1999).}
%
\bibitem{cleo}{K.W.~Edwards {\it et al.} (CLEO Collaboration),
         Phys. Rev. Lett. {\bf86}, 30 (2001).}
%
\bibitem{NIM}{K.~Abe {\it et al.} (Belle Collaboration),
	KEK Progress Report 2000-4 (2000),
	to be published in Nucl. Inst. and Meth. A.}
%
\bibitem{KEKB}{KEKB B Factory Design Report, KEK Report 95-7 (1995),
	unpublished; Y. Funakoshi {\it et al.}, Proc. 2000
        European Particle Accelerator Conference, Vienna (2000).}
%
\bibitem{khh}{K.~Abe~{\it et al.} (Belle Collaboration), BELLE-CONF-0114;
         Submitted as a contribution paper to LP2001.}
%
\bibitem{VCal}{D.M.~Asner {\it et al.} (CLEO Collaboration),
        Phys. Rev. D {\bf 53}, 1039 (1996).}
%
\bibitem{PDG}{D.E.~Groom {\it et al.} (Particle Data Group), Eur. Phys. J.
              C {\bf15}, 1 (2000).}
%
%\bibitem{ARGUS}{ARGUS Collaboration, H.~Albrecht {\it et al.}, Phys. Lett. 
%               B {\bf229}, 304 (1989).}
%


\end{thebibliography}
\end{document}